\documentstyle[12pt,epsf]{article}

\begin{document}
\renewcommand{\thepage}{}

\begin{titlepage}


\title{
\hfill
\parbox{5cm}{\normalsize 
\normalsize UT-776\\
}\\
\vspace{2ex}\large
Non-perturbative evaluation of the effective potential
of $\lambda\phi^4$ theory at finite temperature
under the super-daisy approximation
\vspace{2ex}}
\author{\large
Jiro Arafune\thanks{e-mail address:
 {\tt arafune@icrhp3.icrr.u-tokyo.ac.jp}},\ 
Kenzo Ogure\thanks{e-mail address:
 {\tt ogure@icrhp3.icrr.u-tokyo.ac.jp}}\\
 {\it Institute for Cosmic Ray Research,
   University of Tokyo, Midori-cho,}\\
   {\it   Tanashi, Tokyo 188, Japan}\\
and\\
Joe Sato\thanks{e-mail address:
 {\tt joe@hep-th.phys.s.u-tokyo.ac.jp}}\\
 {\it Department of Physics, School of Science,
   University of Tokyo,}\\
   {\it   Tokyo 113, Japan}}
\date{\today}

\maketitle

\begin{abstract}
    We calculate the effective potential of $\lambda\phi^4$ theory at
    finite temperature under the super-daisy approximation, after
    expressing its derivative with respect to mass square in terms of
    the full propagator.This expression becomes the self-consistent
    equation for the derivative of the effective potential.We find the
    phase transition is first order with this approximation.We compare
    our result with others.

\end{abstract}

\end{titlepage}

\newpage
\renewcommand{\thepage}{\arabic{page}}

\section{introduction}

The symmetry restoration, or the phase transition, at high temperature
is regarded today to play an important role in particle physics and
cosmology\cite{KL}. The investigation 
of the phase transition, however, is found difficult quite often
because of the unreliability of the
perturbation theory at high temperature. In case of the electroweak
phase transition, which may be important for the
baryogenesis\cite{Kux2}, for example, such a breakdown of the ordinary 
perturbative expansion occurs for the Higgs boson mass $m_H$ to be
equal to or larger than the weak boson mass $M_W$\cite{Arn2}. The
experimental evidences\cite{PDG}, in fact, suggest that $m_H$ is not
smaller than $M_W$, indicating that we cannot examine the electroweak
phase transition by the perturbation theory.

The difficulty is caused by the Bose condensation in the thermal bath, 
and is a well-known feature inherent to the finite temperature field
theory\cite{Sha,Arn3}.
The traditional method to improve the perturbation theory is to resum
the daisy diagrams\cite{Arn2,DJ,Wei,Fen,Kap,Bel,Car,Tak}
and use the mass $m_T^2$, the sum of the thermal mass and the zero temperature
mass, instead of the latter only. 
But this improvement is not sufficient.
When one uses the loop expansion, 
one can estimate how small a contribution the next order loop has
in comparison with the previous order loop\cite{Arn2,Fen}. 
The ratio of these is called a loop expansion parameter.
After resumming the daisy diagrams,
we find the loop expansion parameter is $\frac{\lambda T}{m_T}$
in the $\lambda\phi^4$ theory for the temperature $T$. It is $O(1)$ 
near the critical temperature and the analysis with the daisy diagram
indicates the phase transition to be of first order 
though 
it is well known to be of second order\cite{Arn2,Pes}.
This shows the perturbation theory breaks down near the critical
temperature and is not reliable. Many non-purturbative approaches,
such as lattice Monte Carlo \cite{KLRS},
epsilon expansion\cite{AY}, CJT method\cite{CJT,AC,AP},
effective three dimensional theory\cite{Kaj}, 
gap equation method\cite{Buc} 
etc., have been made to investigate the phase transition at high temperature.
However, they have not reached a consensus on what parameter region of
$m_H$ and $M_W$the phase transition is of first order.

     Recently, two of the present authors(K.O. and J.S.) proposed the new non-perturbative method to calculate
the effective potential $V$\cite{A}.
We express its derivative with respect
to the mass square, $\partial V/\partial m^2$, in term of the full propagator.
We calculate the effective potential 
by integrating the derivative of the effective potential 
with an initial condition at $m^2=m_0^2$ given in the region where the
perturbation theory is reliable:
\begin{eqnarray}
V(m^2=-\mu^2)=V(m^2=m_0^2)+\int_{m_0^2}^{-\mu^2}\frac{\partial
V}{\partial m^2} d m^2,
    \label{eff}
\end{eqnarray}
where $-\mu^2$ is the mass parameter of the theory and
 $m_0$ is the mass scale at which we give
the initial condition $V(m^2=m_0^2)$.
In principle we can get the effective potential non-perturbatively
by this procedure.
The main problem of this method is how to approximate 
the full propagator in it. In the previous paper we ignored the momentum 
dependence of the self energy graphs and replaced it by the second
derivative of the effective potential with respect to the expectation
value of the field.
In this paper we try another approximation of the full propagator.
We can sum up all the super-daisy diagram contributions correctly in
this method\footnote{
One can also do this using CJT method\cite{CJT}. We will mention this in
Sec.\ref{sum}.
}.
We evaluate the effective potential numerically with this 
approximation.

The paper is organised as follows. In Sec.2 we discuss 
the structure of the evolution equation that we have to solve.
In Sec.3 we introduce the new
approximation. In Sec.4 we show the numerical result. In
Sec.5 we summarise and discuss our result, comparing our method 
with the others.

\section{The structure of the evolution equation}

     In this section we discuss the evolution      
of the effective potential with respect to the mass.
We consider the $\lambda\phi^{4}$ theory
which is defined by the Lagrangian density
\begin{equation}
     {\cal L}_{E}=-\frac{1}{2}
     \left(\frac{\partial \phi}{\partial \tau}\right)^{2}
     -\frac{1}{2}(\mbox{\boldmath $\nabla$} \phi)^{2}
     +\frac{1}{2}\mu^{2}\phi^{2}
     -\frac{\lambda}{4!}\phi^{4}
     +J\phi + c.t.
\label{lag}
\end{equation}
In the previous paper\cite{A} we derived
the evolution equation of the effective potential,
\begin{eqnarray}
     \frac{\partial V}{\partial m^{2}}&=&
     \frac{1}{2}\bar{\phi}^{2}+\frac{1}{4\pi i}
     \int^{+i\infty +\epsilon}_{-i\infty +\epsilon}dp_{0}
     \int\frac{d^{3}\mbox{\boldmath $p$}}{(2\pi)^{3}}
     \frac{1}{-p_{0}^{2}+\mbox{\boldmath $p$}^{2}+m^{2}+\frac{\lambda}{2}
     \bar{\phi}^{2}+\Pi}\frac{1}{e^{\beta p_{0}}-1}\nonumber \\
     &&+\frac{1}{4\pi i}\int^{+i\infty}_{-i\infty}dp_{0}
     \int\frac{d^{3}\mbox{\boldmath $p$}}{(2\pi)^{3}}
     \frac{1}{-p_{0}^{2}+\mbox{\boldmath $p$}^{2}+m^{2}
     +\frac{\lambda}{2}\bar{\phi}^{2}+\Pi}\nonumber \\
     &&+(Z_{m}Z_{\phi}-1)\biggl[\frac{1}{2}\bar{\phi}^{2}
     +\frac{1}{4\pi i}
     \int^{+i\infty +\epsilon}_{-i\infty +\epsilon}dp_{0}
     \int\frac{d^{3}\mbox{\boldmath $p$}}{(2\pi)^{3}}
     \frac{1}{-p_{0}^{2}+\mbox{\boldmath $p$}^{2}+m^{2}+\frac{\lambda}{2}
     \bar{\phi}^{2}+\Pi}\frac{1}{e^{\beta p_{0}}-1}\nonumber \\
     &&+\frac{1}{4\pi i}\int^{+i\infty}_{-i\infty}dp_{0}
     \int\frac{d^{3}\mbox{\boldmath $p$}}{(2\pi)^{3}}
     \frac{1}{-p_{0}^{2}+\mbox{\boldmath $p$}^{2}+m^{2}
     +\frac{\lambda}{2}\bar{\phi}^{2}+\Pi}\biggr].
    \label{evo}    
\end{eqnarray}
where $\Pi=\Pi(\mbox{\boldmath $p$}^{2},
-p_{0}^{2},\bar{\phi},m^{2},\tau)$ 
is the sum of all the one particle irreducible(1PI) self energy graphs.

The first term of Eq. ($\ref{evo}$)
is the simple background field contribution.
The second term gives the finite temperature contribution.
The third term expresses the effect which remains finite at zero temperature.
The last term is the counter term contribution.

\begin{figure}
\unitlength=1cm
\begin{picture}(16,3.5)
\unitlength=1mm
\put(10,6){\Huge $\Pi$}
\centerline{
\epsfxsize=13cm
\epsfbox{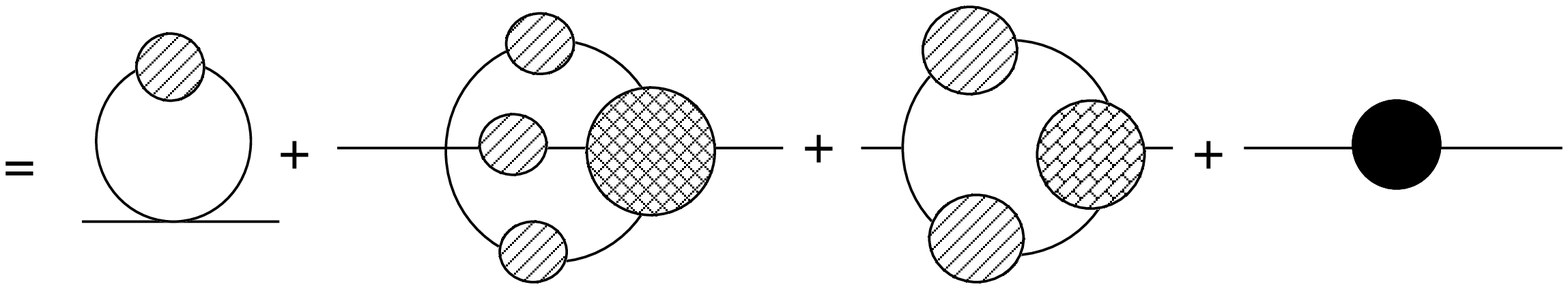}
}
\end{picture}
\caption{1 PI self energy graphs of $\Pi$.
The line with striped blob represents the full propagator.
The circle with a net expresses the full 4-point
vertex(the third term in RHS). The circle
with bricks represents the full 3-point vertex(the
forth term in RHS). The last with a black blob term of RHS represents the counter term.
}
\label{fig2}
\end{figure}

\begin{figure}
\unitlength=1cm
\begin{picture}(16,6)
\unitlength=1mm
\centerline{
\epsfxsize=13cm
\epsfbox{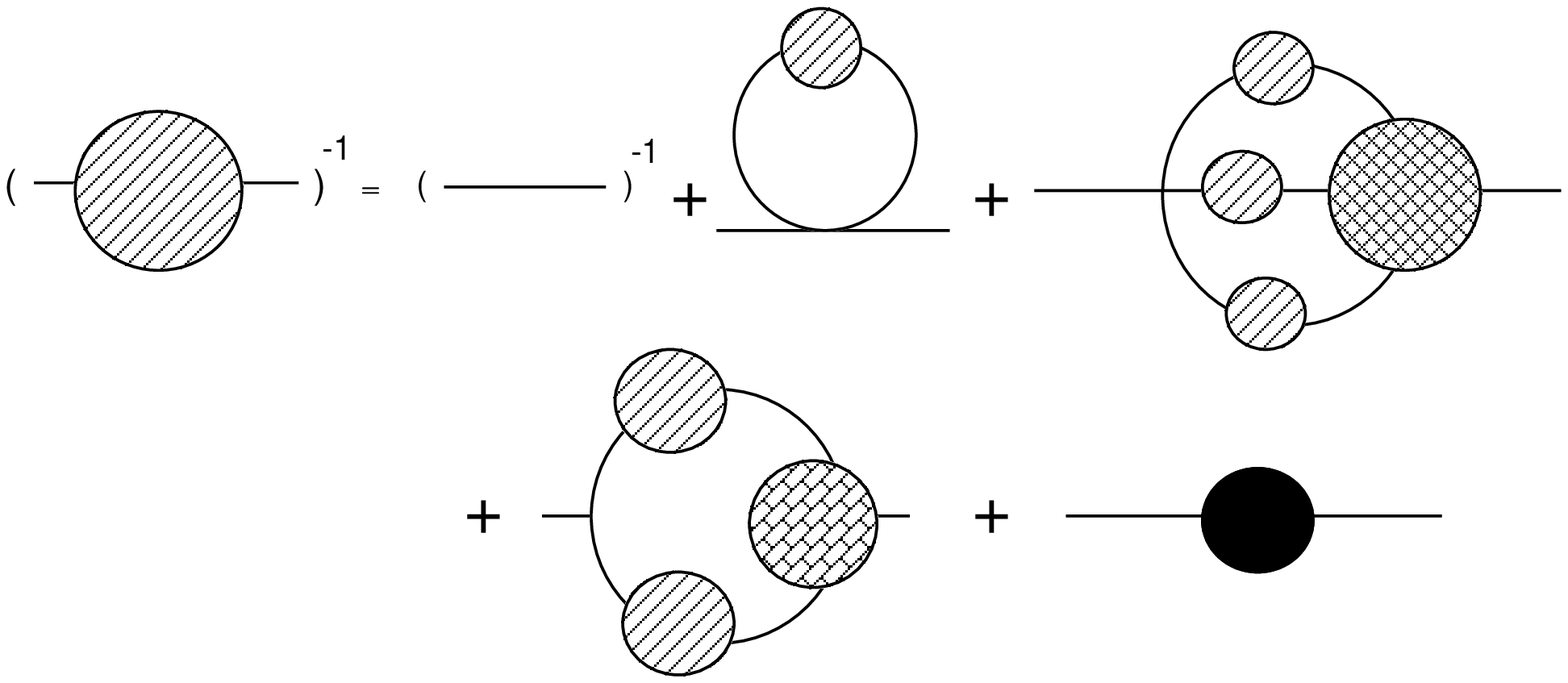}
}
\end{picture}
\caption{Schwinger - Dyson equation for the full propagator
in $\lambda\phi^4$ theory. Symbols are the same as in Fig.\ref{fig2}.
}
\label{fig1}
\end{figure}

\section{Super-Daisy approximation}

\begin{figure}
\unitlength=1cm
\begin{picture}(16,6.5)
\unitlength=1mm
\centerline{
\epsfxsize=9cm
\epsfbox{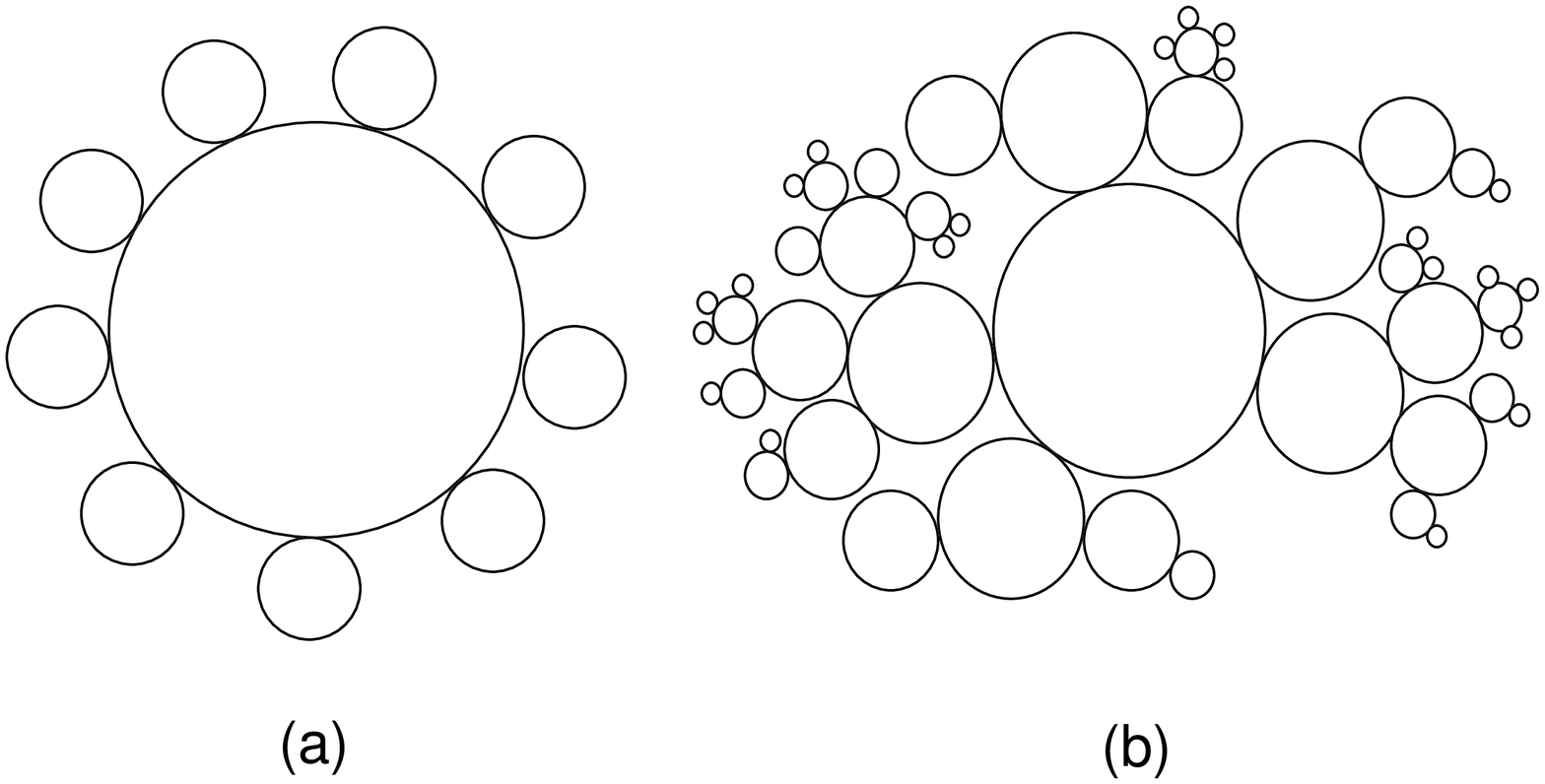}
}
\end{picture}
\caption{Examples of super-daisy diagrams. The graph (a) is often
called merely a daisy diagram.
}
\label{fig3}
\end{figure}

The main
problem of our method is how to approximate $\Pi$ in the
Eq.(\ref{evo}).
The function $\Pi$ is expressed by the full propagator, the full three and
the full four point
functions (Fig.\ref{fig2}) according to the Schwinger - Dyson equation.
The Schwinger - Dyson equation expresses the inverse of
the full propagator by the relation expressed by Fig.\ref{fig1}.
In the right hand side of Fig.\ref{fig2},
the first and the forth terms are independent of the external momentum 
and the second and
the third terms depend on them.
In this paper we assume the simplest case that the momentum
independent term (the first term in Fig.\ref{fig2}) is dominant. 
This corresponds to the 
well-known super-daisy approximation\cite{DJ} which sums up all 
the graphs like
Fig.\ref{fig3}. We stress that
the effective potential
consists of all the super-daisy
diagrams without over counting by this approximation.

\begin{figure}
\unitlength=1cm
\begin{picture}(16,7)
\unitlength=1mm
\put(20,46){\Huge${\partial V\over \partial m^2}$}
\put(50,46){\Huge${1\over 2}\ \bar\phi^2$}
\put(25,9){\Huge$\Pi$}
\put(95,9){\Huge$\lambda$}
\put(104,9){\Huge$\times$}
\centerline{
\epsfxsize=13cm
\epsfbox{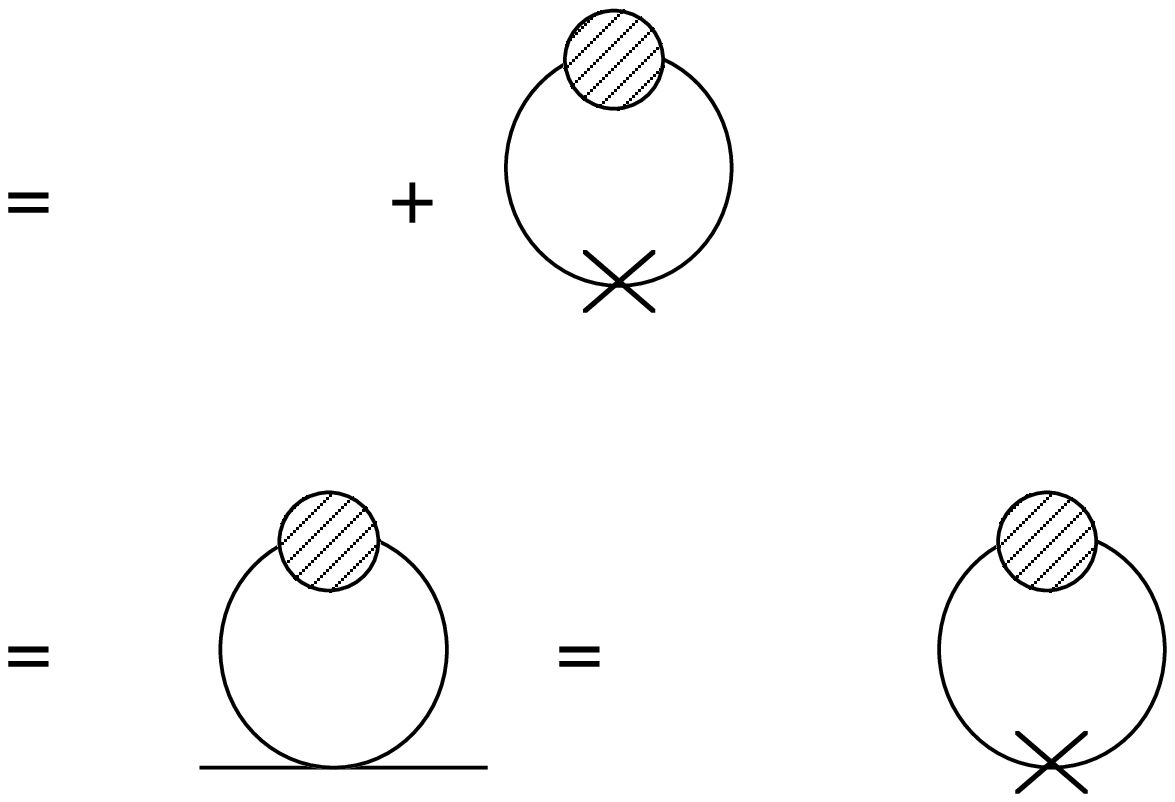}
}
\end{picture}
\caption{The relation between the self energy $\Pi$ and
the derivative of the effective potential $V$ with respect to
the mass square. See the first two term of Eq.(\ref{evo}).
}
\label{fig4}
\end{figure}

In this approximation we see the following relation(Fig.\ref{fig4}),
\begin{eqnarray}
\Pi = \lambda (\frac{\partial V}{\partial m^2}-\frac{1}{2} {\bar \phi^2})
    \label{pi}.    
\end{eqnarray}
In the following we ignore the third and forth terms in
Eq.$(\ref{evo})$, considering 
that only the first two terms are important to discuss the phase
transition\cite{A}.
Due to this approximation, we neglect the loop contribution remaining
finite at $T=0$.
Using relation (\ref{pi}) and integrating Eq.(\ref{evo}) over $p_0$
\cite{Kap,Bel,Kis}, we obtain the approximate evolution 
equation,
\begin{eqnarray}
     \frac{\partial V}{\partial m^{2}}=
     \frac{1}{2}\bar{\phi}^{2}+\frac{1}{2\pi^{2}}
     \int^{\infty}_{0}dp\frac{p^2}{\displaystyle 2\sqrt{p^{2}
     +m^2+\lambda\frac{\partial V}{\partial m^2}}}
     \frac{1}{\displaystyle \exp\left(\frac{1}{T}\sqrt{p^{2}
     +m^2+\lambda\frac{\partial V}{\partial m^2}}\right)-1}.
    \label{napp}
\end{eqnarray}

\section{Numerical result}
\label{num}
In the previous section we have obtained the evolution equation for 
$\frac{\partial V}{\partial m^{2}}$ under the super-daisy approximation.
In this section we explain details of the numerical calculation and
show its result.

\subsection{Details of the numerical calculation}

We can calculate the effective potential, by solving the Eq.
($\ref{napp}$) and integrating it according to the Eq.
(\ref{eff}) numerically.

In the Eq. (\ref{eff})
we set the initial scale $m_0$ as large as $T$ so
that we can evaluate the effective potential very well
by the loop expansion.
We use the same initial condition as was used in the previous paper\cite{A}, 
\begin{eqnarray}
     V(m^2=m_0^2)&=&\frac{1}{2}m_0^{2}\bar{\phi}^{2}
     +\frac{\lambda}{4!}\bar{\phi}^{4}
     +\frac{T^{2}}{2\pi^{2}}
     \int^{\infty}_{0}dp p^{2}\log \left[
     1-\exp\left(-\frac{1}{T}\sqrt{p^{2}+m_0^{2}+
     \frac{\lambda}{2}\bar\phi^{2}}\right)\right].\nonumber\\
    \label{int}
\end{eqnarray}

The integral in the evolution Eq. (\ref{napp})
is well defined when the effective mass square, 
$m^2+\lambda\frac{\partial V}{\partial m^2}$, 
is real and positive. 
Below the critical temperature, however, 
the effective mass square can be negative or complex. We
needs analytic continuation of the integral of the Eq.(\ref{napp}). 
This can be done by rewriting the Eq.(\ref{napp}) as in the previous
paper\cite{A}, 
\begin{eqnarray}
     \frac{\partial V}{\partial m^{2}}=
     \frac{1}{2}\bar\phi^{2}+\frac{T^2}{4\pi^{2}}
     \int^{\infty}_{0}dz
     \frac{\sqrt{z(z+2M)}}{e^{z+M}-1},
    \label{cont}
\end{eqnarray}
where $M={1\over T}\sqrt{m^2+\lambda \frac{\partial V}{\partial m^2}}$.
To find $\partial V/\partial m^2$ we 
solve Eq.(\ref{cont}).

As is well-known, the effective potential becomes complex at small 
${\bar\phi}$ below the critical temperature.
It indicates instability
of the state and the imaginary part of the effective potential
is interpreted to be related with a decay rate of
the state\cite{WW}.
The imaginary part arises from the integral of the  evolution equation in
our method (see Eq.(\ref{cont})).
It is natural to suppose that the imaginary part of the effective
potential is negative to interpret it with decay rate.
In order that the imaginary part of the effective 
potential be negative, 
the imaginary part of $\frac{\partial V}{\partial m^2}$ must be
positive(see Eq.(\ref{eff})). 
We find that there are two solutions for the Eq. (\ref{cont}) and 
that their imaginary part has the opposite sign.
We choose the solution of
$\frac{\partial V}{\partial m^2}$ with positive imaginary part.

\subsection{Result}

\begin{figure}
\unitlength=1cm
\begin{picture}(14,10)
\unitlength=1mm
\put(95,40){$t=4.9288$}
\put(94,15){$t=4.9268$}
\put(100,6){$t=4.9248$}
\put(0,100){${\rm Re}{V(\bar\phi,T)\over\mu^4}$}
\put(135,2){$\bar\phi/\mu$}
\centerline{
\epsfxsize=10cm
\includegraphics{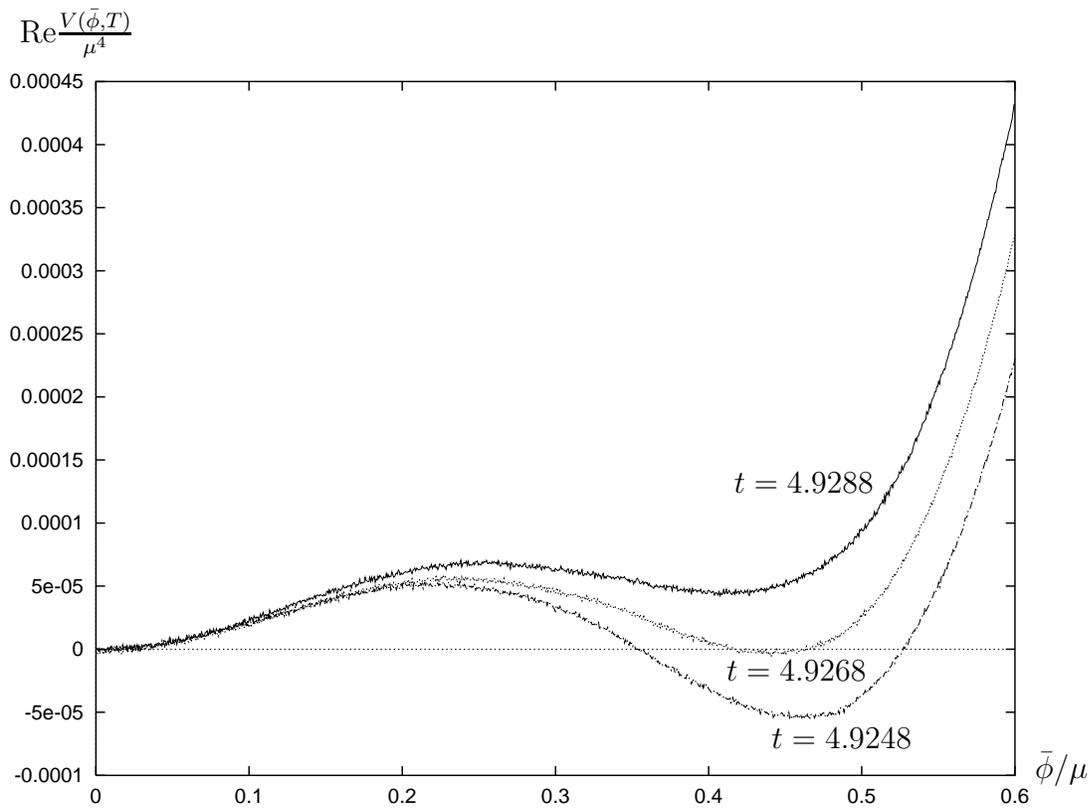}
}
\end{picture}
\caption{Real part of the effective potential at $T=\mu t$.
We see a small barrier to indicate a first order phase transition.
}
\label{fig5}
\end{figure}

\begin{figure}
\unitlength=1cm
\begin{picture}(14,10)
\unitlength=1mm
\put(5,100){$\bar\phi/\mu$}
\put(135,2){$t=T/\mu$}
\centerline{
\epsfxsize=13cm
\includegraphics{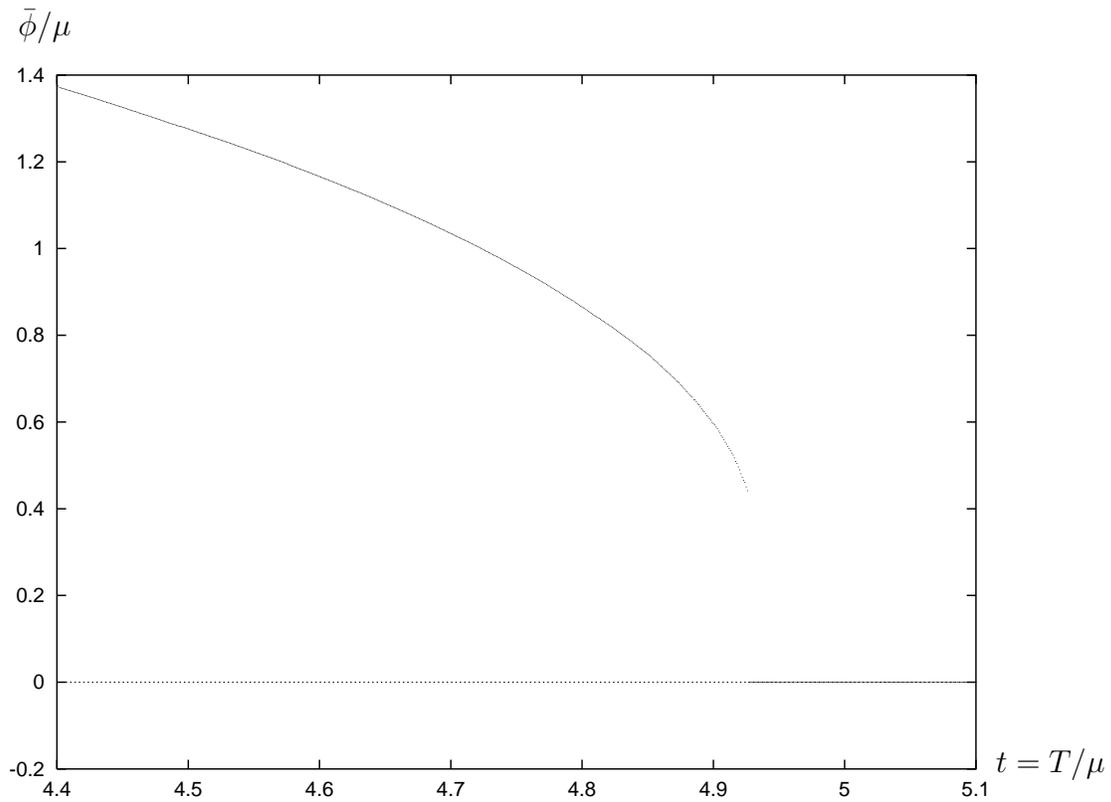}
}
\end{picture}
\caption{
The field expectation value near the critical temperature
at the stable point. We see a finite jump of $\bar\phi/\mu$ at 
$t=4.93\mu$ to indicate a first order phase transition.
}
\label{fig6}
\end{figure}

\begin{figure}
\unitlength=1cm
\begin{picture}(14,10)
\unitlength=1mm
\put(5,100){${\rm Im}{V(\bar\phi,T)\over\mu^4}$}
\put(138,2){$\bar\phi/\mu$}
\centerline{
\epsfxsize=13cm
\includegraphics{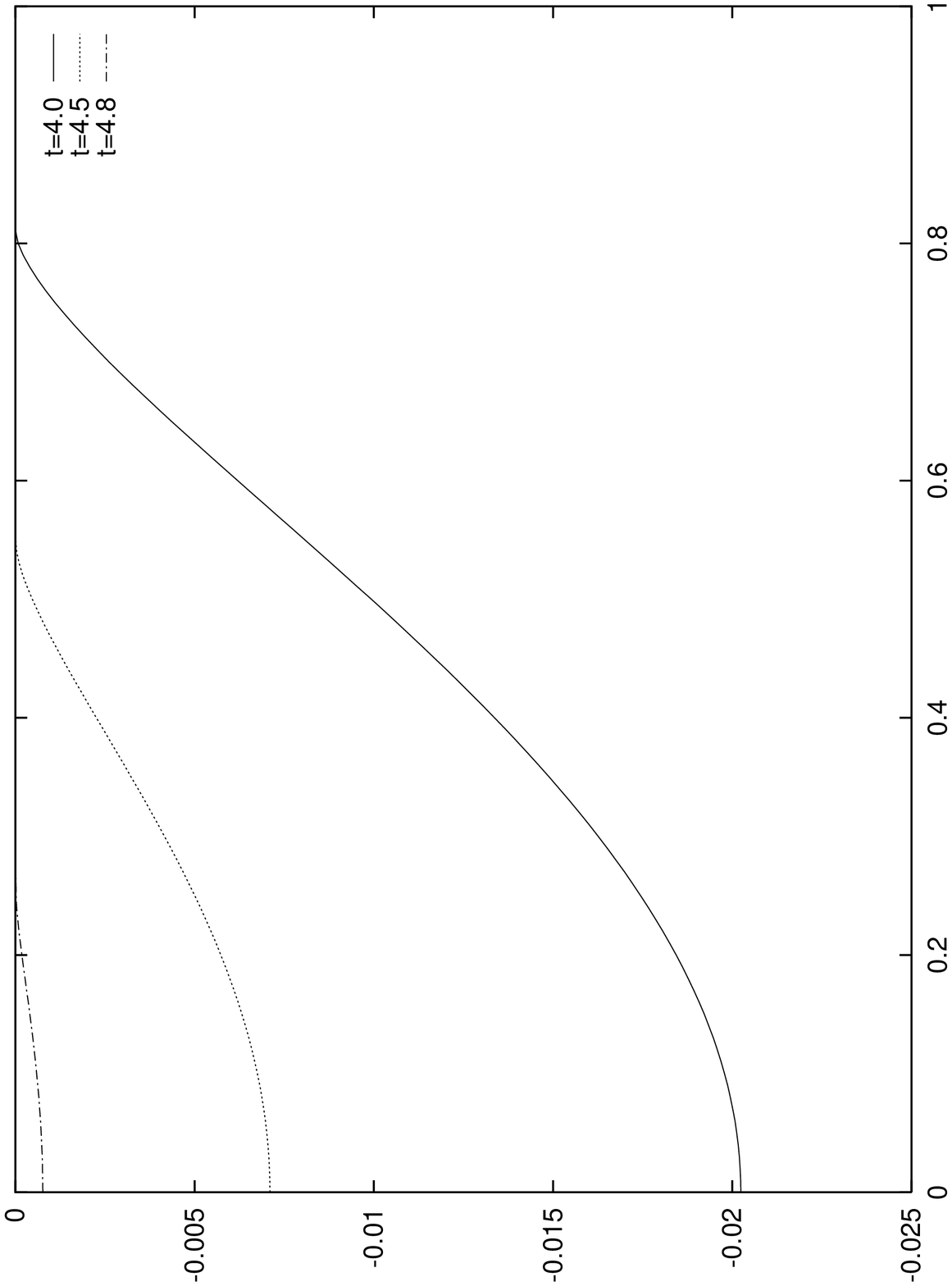}
}
\end{picture}
\caption{
Imaginary part of the effective potential at $T=\mu t$.
We see the smaller $\bar\phi/\mu$ is, the larger the imaginary
part is.
}    
\label{fig7}
\end{figure}

Let us explain the numerical result.
For the graphs Fig.$(\ref{fig5})\sim(\ref{fig7})$ we set\footnote{
We varied the value of $\lambda$ and found no qualitative change.
} the four-point coupling $\lambda=1$. 

We show the real part of the effective potential near the critical
temperature in Fig.\ref{fig5}. We see a small barrier between the
symmetric vacuum ($\bar\phi=0$) and the symmetry-broken vacuum
($\bar\phi\ne0$),which clearly shows a first order phase transition
to occur. 

A first order phase transition is also indicated by the behaviour
of the field expectation value
at the stable point, since Fig.\ref{fig6} shows 
there is a finite jump of it.

We get the negative imaginary part of the effective potential 
below the critical temperature and
show it in Fig.\ref{fig7}.
We can see that the magnitude of the imaginary part increases as the
field expectation value decreases.
It expresses that the closer the field expectation value
to the origin is,
the less stable the state below the critical temperature is.

In the above we used the initial condition at $m=m_0=T$.
It is required $m_0$ be the same order as the temperature
in order that the initial condition can be evaluated 
by the perturbation.
We have calculated the effective potential with other initial conditions
 $m_0=2T$ and $\frac{T}{2}$ and do not have found any
appreciable change of the effective potential.
It ensures the consistency of our calculation.

\section{Summary and discussion}
\label{sum}
In this paper we have given a new method to calculate the effective
potential of $\lambda\phi^4$ theory under the super-daisy
approximation without over counting.  
We have numerically evaluated a real part and a imaginary part of the
effective potential without using high temperature expansion. 
The real part indicates a first order phase transition 
though the true order should be second. 
The imaginary part indicates the instability is larger for the smaller
expectation value of the field below the critical temperature.

\begin{figure}
\unitlength=1cm
\begin{picture}(16,4)
\unitlength=1mm
\centerline{
\epsfxsize=13cm
\epsfbox{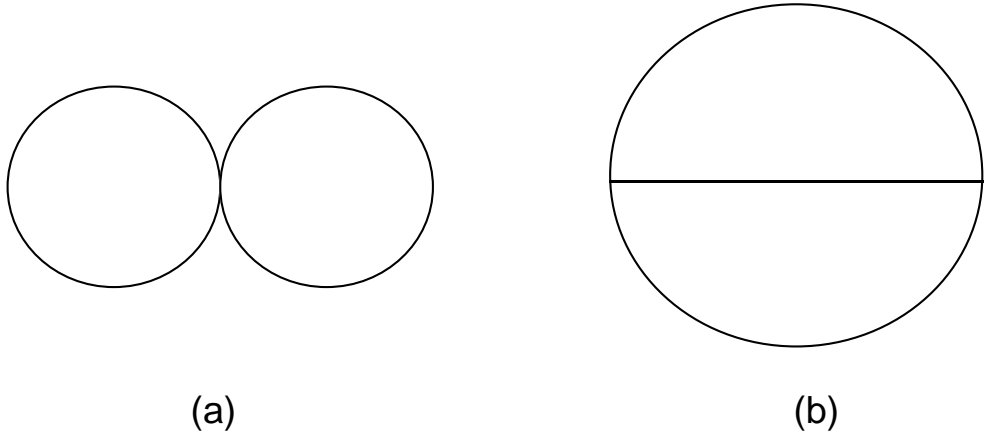}
}
\end{picture}
\caption{
Two loop diagrams that contribute to the effective potential.
(a) is daisy-like and (b) is not daisy-like.
}
\label{fig8}
\end{figure}

Now we compare our method with other approaches.
First we compare it with CJT method\cite{CJT} which can also gather up the
super-daisy graphs without over counting.
It carried out by truncating the CJT expansion at $O(\lambda)$. 
A first order phase transition is indicated by the calculation of
this model with this method as well as the high temperature expansion\cite{AP}.

Second we compare with daisy-improved perturbation theory.
In this method the effective potential indicates a first order phase 
transition at one loop level\cite{Tak}. There are two graphs at two loop
level(see Fig.\ref{fig8}).
If one includes the 
contribution of Fig.\ref{fig8}(a) alone\cite{Tak},
the phase transition is still of first order\footnote{
In ref.\cite{Tak} only Fig.\ref{fig8}(a) was included.
}.
A second order phase transition is indicated when the
contributions of both Fig.\ref{fig8}(a) and Fig.\ref{fig8}(b) 
are included\footnote{
In ref.\cite{Arn2} they calculated the effective potential including both of Fig.\ref{fig8}(a) and Fig.\ref{fig8}(b).
They did not, however, calculate the expectation value of the field
nor argue on the order of the phase transition with their 2-loop result.
We calculated the expectation value of the field
using their effective potential
and found the phase transition second to be of second order.
}. 
This indicates that
the contribution of Fig.\ref{fig8}(b), which is not daisy-like, seems to play
the important
role to the effective potential.

Finally we compare our result with the approximation of the previous paper.
In the previous paper we get the effective potential that indicates
a second order phase transition.
The contributions of the second and the third terms of Fig.\ref{fig2} with zero
external momentum are taken into account by the previous approximation.
The difference of the two
approximations is whether we include the contributions of the 
daisy-like diagrams only or not.
This also indicates that non-daisy-like diagrams are important for the 
second order phase transition to be derived correctly.

According to the above comparisons, the diagrams which are not daisy
like seem to give important contributions.
We have to take into account the contribution of the second and the
third terms in Fig.\ref{fig2}. 
Their momentum dependence, especially
the dependence on the small momentum, will be important in Eq.(\ref{evo})
since the infrared behaviour of the self energy
plays a very important role for a second order phase
transition\cite{Wilson}.
This method with $\frac{\partial V}{\partial m^2}$ seems attractive because it may bring a chance to step in the
region where the perturbation theory breaks down.

We finally express our 
thanks to A. Niegawa and
T. Inagaki for valuable discussions and 
communications.

\end{document}